\documentclass[aps,preprintnumbers,twocolumn]{revtex4}
\usepackage{amsfonts,amssymb,amsmath}            % for math symbols.
\usepackage[]{graphics,graphicx,epsfig}            % for graphics figures.
\usepackage{amsthm}
\usepackage{subfigure}                              
\def\identity{\leavevmode\hbox{\small1\kern-3.8pt\normalsize1}}

\newcommand{\ket}[1]{\left | #1 \right\rangle}
\begin{document}

\title{Error Correcting the Control Unit in Global Control Schemes}
\date{\today}

\author{Alastair \surname{Kay}}
\email[]{alastair.kay@qubit.org} \affiliation{Centre for Quantum
Computation,
             DAMTP,
             Centre for Mathematical Sciences,
             University of Cambridge,
             Wilberforce Road,
             Cambridge CB3 0WA, UK}
             
\begin{abstract}
Recent studies of globally controlled structures have culminated in a theoretical demonstration that fault-tolerant quantum computation can be carried out on a one--dimensional chain with control over two global fields only. This required some patterns of classical states to localise operations, which were stabilised with the Zeno effect. However, it is impossible to achieve perfect stabilisation using this method, so error correction of these states is desirable, and is the focus this paper.
\end{abstract}

\maketitle
Physical implementations of quantum computers typically require incredibly precise control over every single element of the system. Such precision is a major barrier to the realisation of a quantum computing device. In some, otherwise promising, systems, such precision is simply not possible. As a result, other paradigms for quantum computation have been proposed. For example, cluster state computation reduces the problem of arbitrary two--qubit and one--qubit gates to one--qubit operations and a single mass entangling operation \cite{cluster1}. A further approach, that of global control, has also been the subject of recent investigation \cite{Benjamin:2001a,Benjamin:2002a,Benjamin:2002b,Benjamin:2003b}, with several practical implementations having been proposed \cite{BenjaminBose1,Kay:2004a,Zoller:1,Kay:2005a}.

In a global control scenario, after initialisation of the system, we assume that the only control that is available governs global fields. For theoretical investigations, we typically restrict ourselves to a one--dimensional chain of qubits, as this is minimal. Global control pulses can then be considered as issuing commands such as ``Apply a Hadamard gate to alternate qubits''. Even under such restrictions, it has been demonstrated that error correction, and fault-tolerance, can be implemented efficiently \cite{Benjamin:2003b}. In that paper, the intention was to provide a proof of principal. As such, the authors concentrated simply on correcting the computational qubits in the system. However, one of the costs of a global control scheme is an encoding of qubits which means that there are several blank qubits associated with every computational qubit. These were not corrected in the scheme of \cite{Benjamin:2003b}, and were simply stabilised with the Zeno effect. Furthermore, the way that the effect of the global fields are localised is to introduce a localised imperfection in the system (known as the Control Unit, CU). This imperfection can be moved around the system so that it can interact with any of the computational qubits. The stability of the CU is, of course, critical. If it were to disappear, the whole computation would stop.

In this paper, we concentrate on error correcting the CU and the buffer qubits (all those that aren't computational qubits). From the number of different encodings that Benjamin has proposed for 1D global control schemes, we shall select \cite{Benjamin:2002a}. This encoding lends itself particularly well to error correction because the computational qubits are single spins, and not encoded. Such a choice automatically avoids a particular concern of other encodings, that of ensuring the state of the qubits remain in the computational basis. There are two key ingredients to this error correcting scheme. Firstly, we must realise that all the buffer qubits are in classical states. This means that any regular patterns of qubits can simply be reset to the value which we know they should be. Secondly, for structures such as the CUs, we will introduce a redundancy which will allow a typical quantum error correction scheme \cite{nielsen} with adaptation to allow feedback of the error syndrome in an algorithmic manner \cite{Benjamin:2003b}. We will demonstrate how to perform standard elements of a computation and error correction in the presence of this redundancy.

\begin{figure}[!t]
\begin{center}
\includegraphics[width=0.45\textwidth]{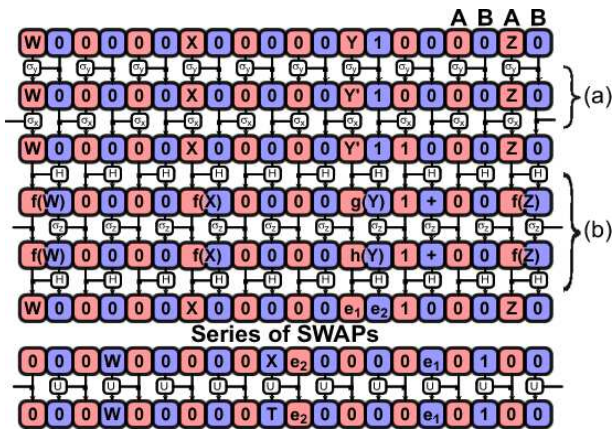}
\caption{Method for creating a two--qubit gate between distant qubits. The sequence up to the SWAPs places the information of the qubit Y onto the CU ($\ket{e_1e_2}=\alpha\ket{10}-\beta\ket{01}$ where $\ket{Y}=\alpha\ket{0}+\beta\ket{1}$), and subsequently this is used to act upon the target qubit, X. After the controlled-$U$ step, all previous steps must be undone to disentangle the pointer. Part (a) translates the CU into a unique local pattern in the $A$ qubits. Part (b) flips $B$ qubits if they are surrounded by two $\ket{1}$s.}
\label{2_qubit_gate}
\end{center}
\end{figure}

Before presenting the main results of this paper, we briefly review the basic model that we are using \cite{Benjamin:2002a} and its extension to error correction \cite{Benjamin:2003b}. Consider a one--dimensional array of qubits. We shall label them alternately $A$ and $B$, and they are all initialised in the $\ket{0}$ state. There are four basic operations that we allow - single qubit operations (rotations, measurement and a `reset' operation, which sets the qubit to $\ket{0}$ irrespective of its initial state) on either the $A$ qubits or the $B$ qubits and controlled-phase gates between qubits paired up as $AB \;AB \;AB...$ or $BA \;BA \;BA...$. Such procedures are readily composed to give arbitrary controlled-$U$ gates \cite{nielsen}, and also form a physically sensible set of procedures \cite{Kay:2004a}. The Control Unit is created by by flipping one of the $B$ qubits to the $\ket{1}$ state. Such initialisation can be accomplished in a variety of ways as a one-off effect \cite{Vollbrecht, Kay:2005a} The computational qubits are located on every third $A$ qubit. By performing SWAP operations between pairs of qubits, the $B$ qubits can be moved relative to the $A$ qubits i.e. the CU can be positioned next to any computational qubit that we choose. Single--qubit gates are then implemented by performing a controlled-$U$ gate where the $B$ qubits are the controls. This only applies the operation $U$ to the computational qubit adjacent to the CU. Extra buffer qubits (on the $A$s) have been introduced to allow the two--qubit gate, which is illustrated in Fig. \ref{2_qubit_gate}.

In order to perform error correction, we introduce a regular array of CUs, each one associated with a block of $L$ qubits, which corresponds to a single qubit encoded against errors by an error correction code, such as the Shor[[9,1,3]] code. This can then be used to implement the relevant correction algorithm on all the encoded qubits simultaneously. Of course, for a standard computational step we also have to be able to switch back to the situation of having only a single CU. This is achieved by an additional pre-patterning of classical states in regions known as Switching Stations (SS). Each is associated with a single CU. In the case of error correction, these SSs are all prepared in the $\ket{0}$ state, except for one, which is in the $\ket{1}$ state. This bit can then be used as the control bit in a controlled-$U$ gate, effectively deactivating all but one of the control units. More complex patterns of SSs, enabling fault-tolerance, are detailed in \cite{Benjamin:2003b}. These create patterns of CUs every $L^i$ qubits for any $i$ up to a maximum value, by placing the classical label $i-1$ in the relevant SS (the label $i+1$ goes in a SS which would otherwise be labelled $i$).

Let us initially consider resetting the buffer qubits in the system, assuming that the CUs are perfect. This involves two steps, one which corrects those on $A$ qubits and another that corrects those on $B$ qubits. Both of these will occur in the regime where there is one CU activated for every SS. Resetting the $A$ buffers is simple - we just move the CU next to each of them and send out the reset command, much as we would for a single--qubit gate. This requires a fixed number of steps, independent of device size.

To correct the $B$ buffers, we move the CU adjacent to the result qubit in the SS (the one which is used as the control bit), and switch it to the $\ket{1}$ state. We also do that to an adjacent $A$ qubit (i.e. the nearest physical $A$ qubit, not the next computational qubit). This creates a unique local patterning which is used to feed operations back onto the $B$ qubits (see Fig. \ref{2_qubit_gate}). Hence we move all the buffer qubits between these two $\ket{1}$s and reset them. One potential risk is an interaction with the CU when it is adjacent to one of the $\ket{1}$s during the feedback process. Since the mechanism for the reset procedure will depend on the physical implementation, this is difficult to answer, but we can at least say that if we were applying a unitary operation, there is no additional interaction.

One might think that the unique patterning, the same as found in the two--qubit gate, could be used to deactivate the CU, meaning that all the $B$ qubits should be in the $\ket{0}$ state, so we could globally apply the reset procedure to all of them. Unfortunately, this process would also move the imperfections of the $B$ qubits onto the $A$ qubits, and, in turn, move some of the computational qubits onto $B$ qubits (if there are errors on some of the $A$ buffers). This reminds us that if there are imperfections in the $B$ qubits, these will be mapped onto the $A$ qubits, including the computational qubits. However, error correction will correct for these provided they occur sufficiently infrequently.

Having reset the buffer qubits, we need to reset, or otherwise stabilise, the qubits in the SSs and the CUs. We can use the different levels of concatenation in the fault-tolerant scenario to our advantage for performing resets of different organisations of qubits. Consider the scenario where we have a CU enabled for every $L$ SSs. Between their own SSs, there are $L-1$ SSs which all store the value 0 and have deactivated CUs. Thus, we can reset all of these. Note, however, that they can't reset their own states. At higher levels of concatenation, there is a CU for every $L^i$ SSs. Regularly spaced between each of these are $L-1$ SSs which contain the number $i-1$ and deactivated CUs that have not been corrected yet. So, we can move the active CUs along and reset all of these. These continue up the hierarchy of concatenation until we get to a single CU, suitable for performing the algorithm (instead of just performing error correction). However, this single CU has never been corrected and the accuracy of the correction of all the others depends, to some degree, on how well this is maintained in the $\ket{1}$ state. Hence, we need some other way of correcting this CU. In addition, the SSs for the highest level of concatenation cannot efficiently be corrected by this single CU.

Thus, to correct the CUs, we must consider moving to a more typical error correction scenario. Recall that because we are using classical states, the only errors that we have to protect against are bit flips. This can be done simply by repeating the same state three times. By regularly comparing the three, if one has changed, it can subsequently be corrected by setting it equal to the other two. Of course, if two have changed then the ``correction'' goes in the wrong direction. This can be minimised by increasing the number of copies, but there will be a corresponding overhead in the number of steps required.

So, instead of using a single CU for a computation, we shall now use three of them. We do not intend to perform any part of the algorithm with all three CUs present, as this will be less efficient than switching off two of them. We can initially align these with computational qubits. The patterning needs to be chosen with care in order to minimise the complexity of the protocol. We target an operation on a single qubit by applying a controlled-phase gate ($CP$) with each of the three CUs, separated by the single qubit rotations $U_1$, $U_2$, $U_3$ and $U_4$ applied to all the $A$ qubits. The resulting evolution on the targeted qubit is
$$
U_1\sigma_zU_2\sigma_zU_3\sigma_zU_4.
$$
Any qubits that are far enough away from the CUs will not be affected by the $CP$s, and hence will be subject to the evolution $U_1U_2U_3U_4$, which we select to be the identity transformation i.e. $U_4=U_3^\dagger U_2^\dagger U_1^\dagger$. We need to create sequences such that we can apply an arbitrary rotation to the qubit we want, but do nothing to any of the other qubits. This necessarily includes qubits that get affected by one or two of the $CP$s. The first step is to select a patterning of the CUs such that no qubit ever experiences 2 $CP$s (unless an error has occurred). The simplest example is to align the CUs with qubits $q_1$, $q_2$ and $q_4$ (note the gap, $q_3$). 

In the fault-tolerant scenario presented in \cite{Benjamin:2003b}, the most efficient way of introducing these three CUs is by relabelling the switching stations using the following protocol. If the label was non-zero, add 1 to the value (these correspond to the qubits $q_1$ for each block). Label the zero-valued SSs that are 1 and 3 SSs away from these relabelled SSs with the number 1 (corresponding to qubits $q_2$ and $q_4$). So, by deactivating all CUs in regions controlled by SSs with 0 labels, we get the regular patterning that is required, and we can still access the parallelism required for fault-tolerance. In fact, by repeating this procedure at every level of concatenation, this enables fault-tolerance of the CUs at a cost of a single extra bit in each SS. We will not concentrate on this, as it is a simple generalisation of the ideas demonstrated here.

The 1-2-4 arrangement of CUs can still cause unavoidable applications of single $CP$ gates,
%In the full error correction scenario, this means that we align them with switching stations 1, 2 and 4, where we only count the switching stations that store the number $p-1$ (where we have $p$ levels of concatenation). These would be uniquely marked by changing the value of those SSs to the number $p$. This patterning still causes unavoidable single applications of $CP$,
\begin{eqnarray}
&U_1\sigma_zU_1^\dagger&	\nonumber\\
&U_1U_2\sigma_zU_2^\dagger U_1^\dagger&	\nonumber\\
&U_1U_2U_3\sigma_zU_3^\dagger U_2^\dagger U_1^\dagger.&	\nonumber
\end{eqnarray}
Clearly, if we apply the sequence {\em{twice}}, all of these will be removed, while the targeted qubit experiences the evolution
$$
U_1\sigma_zU_2\sigma_zU_3\sigma_zU_3^\dagger U_2^\dagger \sigma_zU_2\sigma_zU_3\sigma_zU_3^\dagger U_2^\dagger U_1^\dagger
$$
which, given that we have a free choice of $U_1$, $U_2$ and $U_3$, must contain sufficient freedom to create any single qubit rotation that we desire (up to a global phase).

Now that we are using a CU that has some redundancy in it, there is enough information to be able to correct for errors. There are two stages involved in making use of this information. Firstly, we must perform a syndrome extraction, placing information about any errors that have occurred on an ancilla qubit (which would otherwise have been a computational qubit). Secondly, we must feed back from this ancilla to be able to correct the faulty part of the CU.

If one of the CUs suffers a bit flip, then only the targeted qubit will be affected. We can therefore neglect all other qubits, and just concentrate on the ancilla that we will be targeting, and which will initially be in the state $\ket{0}$. For the three bits that can get flipped, the resulting evolution will be one of
\begin{eqnarray}
&U_1U_2\sigma_zU_3\sigma_zU_3^\dagger \sigma_zU_3\sigma_zU_3^\dagger U_2^\dagger U_1^\dagger,&	\nonumber\\
&U_1\sigma_zU_2U_3\sigma_zU_3^\dagger U_2^\dagger\sigma_zU_2U_3\sigma_zU_3^\dagger U_2^\dagger U_1^\dagger,&	\nonumber\\
\text{or}&U_1\sigma_zU_2\sigma_zU_2^\dagger\sigma_zU_2\sigma_zU_2^\dagger U_1^\dagger. &	\nonumber
\end{eqnarray}
Using either $U_2=\identity$ or $U_3=\identity$, the evolution when the CU has no error is $\identity$. The results if there have been errors are shown in Table \ref{tab:syndrome}, where
$$
V_X=U_1\sigma_zU_X\sigma_zU_X^\dagger\sigma_zU_X\sigma_zU_X^\dagger U_1^\dagger.
$$
$X$ is either 2 or 3, where $U_X\neq\identity$. We thus have free choice of $U_1$ and $U_X$ to make this remaining evolution $\sigma_x$ (for example, $U_1=\identity$ and $U_X=e^{-i\sigma_x\pi/8}$), which enables the error syndrome to be placed on the ancilla.

\begin{table}[!t]
\begin{center}
\begin{tabular}{|c|c|c|}
\hline
Error occurs on CU:	& $U_2=\identity$	&	$U_3=\identity$	\\
\hline
none	&	$\identity$	& $\identity$	\\
1	&	$V_3$	&	$\identity$	\\
2	&	$V_3$	&	$V_2$	\\
3	&	$\identity$	&	$V_2$	\\
\hline
\end{tabular}
\caption{The evolution that can occur if an error affects a single CU out of the three. By selecting the single--qubit rotations suitably, syndrome extraction can be performed.}
\label{tab:syndrome}
\end{center}
\end{table}

\begin{figure}[!t]
\begin{center}
\includegraphics[width=0.45\textwidth]{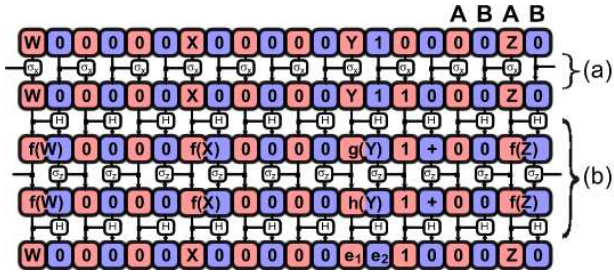}
\caption{Method for feeding back error syndrome to CUs. The specification differs from the first half of Fig. \ref{2_qubit_gate} because the steps in part (a) are now controlled by the 2 CUs that are not being corrected. The first step has also been removed. Part (b) is applied to the CU that is being corrected only.}
\label{2_qubit_gateb}
\end{center}
\end{figure}

How can this information be used to correct the right error? If we apply $V_3=\sigma_x$, i.e. we have set $U_2=\identity$, then the target (ancilla) qubit is flipped if there is an error on either of the first two CUs. We then use the circuit in Fig. \ref{2_qubit_gateb} to feed back the error syndrome from the ancilla to the first CU, where part (a) is controlled by both CUs 2 and 3. If the error occurred on the second CU, then the feedback process won't occur. Therefore, we can correct for a single error on any of the CUs by repeating the process.

The system can therefore be stabilised against independent errors on any of the qubits in the system, a significant improvement over the stability of the error correcting procedure presented in \cite{Benjamin:2003b}. Extension of these ideas enables, by introducing only a single extra bit to the SS, the full fault-tolerance of the CUs, which guarantees fault-tolerance of all the other classical states in the system. These ideas can be applied to any global control scheme, with the additional caveat that we must assume that the system stays in its computational basis.

\end{document}